\documentclass[11pt,twoside]{article}
\usepackage{atmp}

\newcommand{\al}{\alpha'}
\newcommand{\de}{\partial}
\newcommand{\be}{\begin{equation}}
\newcommand{\ba}{\begin{eqnarray}}
\newcommand{\ea}{\end{eqnarray}}
\newcommand{\ee}{\end{equation}}

\newcommand{\f}{\frac}
\newcommand{\s}{\sqrt}

\newcommand{\ap}{\alpha}

\newcommand{\ddd}{\cdot\cdot\cdot}
\newcommand{\no}{\nonumber \\}
\newcommand{\la}{\langle}
\newcommand{\lb}{\rangle}
\newcommand{\ep}{\epsilon}

\copyrightnotice{2003}{7}{369}{379} 
 
\begin{document}
 
\title{Correlators in Timelike Bulk Liouville Theory}
 
\url{hep-th/0303221}		
 
\author{Andrew Strominger and Tadashi Takayanagi} 
\address{Jefferson Physical Laboratory\\
Harvard University\\
Cambridge, MA  02138}
\addressemail{andy@planck.harvard.edu} 
\addressemail{takayana@wigner.harvard.edu}	
 
\markboth{\it Correlators in Timelike Bulk Liouville Theory}
{\it Strominger and Takayanagi}

\begin{abstract}
Liouville theory with a negative norm boson and no screening
charge corresponds to an exact classical solution of closed
bosonic string theory describing time-dependent bulk tachyon
condensation. A simple expression for the two point function is
proposed based on renormalization/analytic continuation of the
known results for the ordinary (positive-norm) Liouville theory.
The expression agrees exactly with the minisuperspace result for
the closed string pair-production rate (which diverges at finite
time). Puzzles concerning the three-point function are presented
and discussed.
\end{abstract}

\cutpage

\section{Introduction}
\setcounter{equation}{0}

The endpoint - if any - of tachyon condensation in closed string
theories with tachyons is a major unsolved problem.  A relatively
complete picture has recently been obtained of the condensation of
certain $localized$ closed string tachyons
\cite{Adams:2001sv,Vafa:2001ra,Harvey:2001wm,David:2001vm}, as
have some insights into the non-localized bulk case
\cite{Gutperle:2001mb}.

 It is
interesting in this regard that there are exact, time dependent
classical solutions of tachyonic string theories describing
homogenous tachyon condensation.  We refer to the corresponding
worldsheet CFT for the case of the bosonic string as the timelike
Liouville theory. It is governed by the the action (suppressing
spatial dimensions and setting $\al =1$) \ba S=\f{1}{4\pi}\int
d^2\sigma  \left( -(\de X^0)^2 +4\pi \mu e^{2\beta X^0}\right),
\label{TLB} \ea which has a negative norm boson and central charge
$c=1-6q^2\ \ (q\equiv\beta-1/\beta)$. This corresponds to a real
dilaton with timelike slope $q$, but we will mainly be interested
in $q=0$ and $\beta=1$.  The potential term in (\ref{TLB}) can be
interpreted a closed string tachyon field which grows
exponentially in time. At early times $X^0\to -\infty$ the tachyon
field is very small and flat space is recovered.

Because this is a time dependent background, there is closed
string pair production. In section 2 we compute the rate of pair
production in a minisuperspace approximation
\cite{Braaten:np,Polchinski:1990mh} and find that it diverges
exponentially (see also \cite{daCunha:2003fm}). This divergence is
much stronger
 than that found in an analogous computation of open
string pair production \cite{St,GS,Maloney:2003ck} 
during open string tachyon
condensation\cite{Sen:2002nu,Sen:2002in} (the related phenomenon
of closed string emission was analyzed
in\cite{Okuda:2002yd,Larsen:2002wc,Lambert:2003zr,Mi}). 
This result implies
that the gas of pair produced closed strings will reach the string
density and string perturbation theory will break down in a time
of order one in string units.

This work began in as an attempt to compute the two and three
point functions of the timelike Liouville theory. This sounds like
a simple task but in fact the analytic continuation involved is
quite subtle. Our strategy will parallel that adopted for the
timelike boundary Liouville theory in \cite{GS}. After some
renormalization and analytic continuation we will produce a
reasonable expression for the exact two point function in section
3 which agrees exactly with the minisuperspace particle production
rate. We also discuss the three point function. We present a
natural analytic continuation procedure which encounters
singularities and yields a puzzling result, not obviously
compatible with conformal invariance, as discussed in section 4.

The basic source of the difficulty is that the action (\ref{TLB})
is not positive definite, and hence does not fully define the
associated functional integral or CFT. It is natural to try to
define the timelike theory\footnote{We assume here that such an exact
CFT indeed exists, although it has yet to be demonstrated.}
 by analytic continuation $\phi=-iX^0$
and $b=i\beta$ from the ordinary spacelike Liouville theory with
the positive definite worldsheet action \ba S=\f{1}{4\pi}\int
d^2\sigma \left( (\de\phi)^2 +4\pi \mu e^{2b \phi }\right), \ea
and background charge $Q=b+1/b$. The central charge  here is
$c=1+6Q^2$ and $Q$ is the (spacelike) slope of the dilaton. The
analytic continuation from real to pure imaginary $b$ however, for
the three point function, leads to an infinite accumulation of
singularities. A similar analytical continuation was discussed in
\cite{Runkel:2001ng}, where it was argued that the $b=i$
singularity is related to the accumulation of minimal models at
$c=1$. It was also discussed in \cite{GS} (whose treatment we
closely follow) in the closely related context of timelike
boundary Liouville theory\cite{FZZ,Teschner:2000md}.\footnote{For
the boundary theory, the singularities accumulate already for the
two point function, while for the bulk theory this difficulty
first arises for the three point function.}

\section{The Minisuperspace Approximation}
\setcounter{equation}{0}

In this section we apply the minisuperspace approximation
\cite{Braaten:np,Polchinski:1990mh} to the time-like bulk
Liouville theory with constant dilaton ($q=0$).\footnote{Recently
the paper \cite{daCunha:2003fm} appeared which also analyzes the
minisuperspace approximation and obtains the same results.} The
analysis here is mathematically quite similar (though the final
conclusion differs) to that given for the boundary tachyon
interaction considered in \cite{St}. We shall accordingly be
brief. In the minisuperspace approximation we retain only the zero
mode $x^0$ of $X^0$. The on-shell condition for the closed string
excitations is then the Klein-Gordon equation with time dependent
mass \ba \Bigl[\f{\de^2}{(\de x^0)^2}+{2\pi\mu} e^{2x^0}+\vec{p}^2
+{2}(N_{L}+N_{R}-2)\Bigr]\psi_{\vec{p}}(x^0)=0. \ea Below we
 define the energy by
$\omega=\s{\vec{p}^2+2(N_{L}+N_{R}-2)}$, and $N_{L,R}$ are the left and
right-moving oscillator contributions.
 $in$ and $out$ wave
function $\psi^{in}_{\vec{p}}(x^0)$ and
$\psi^{out}_{\vec{p}}(x^0)$ asymptotically obeying
$\psi^{in}_{\vec{p}}(x^0)\sim e^{-i\omega
x^0+i\vec{p}\cdot\vec{x}}$ and $\psi^{out}_{\vec{p}}(x^0)\sim
e^{-\f{x^0}{2} -2i\s{\pi\mu}e^{x^0}+i\vec{p}\cdot\vec{x}}$ are
given by Bessel and Hankel functions. This leads to the Bogolubov
coefficients \ba \gamma^{in}_{\vec{p}}
=\f{\beta_{\vec{p}}^{in*}}{\alpha_{\vec{p}}^{in}}=-ie^{-\pi\omega},
\ea where the Bogolubov transformation is \ba
&&\psi^{out}_{\vec{p}}=\alpha_{\vec{p}}\psi^{in}_{\vec{p}}
+\beta_{\vec{p}}\psi^{in*}_{-\vec{p}}\no
&&\psi^{in}_{\vec{p}}=\alpha^*_{\vec{p}}\psi^{out}_{\vec{p}}
-\beta_{\vec{p}}\psi^{out*}_{-\vec{p}}.\no \ea According to the
conjecture of \cite{GS} this is related to the two point function
by \ba |\la e^{-i\omega x^0}e^{-i\omega x^0} \lb|=|\gamma^{in}_{\vec{p}}|
=e^{-\pi\omega}.
\label{tpf} \ea In the following section we shall recover this
result from an exact CFT analysis.

The rate of closed string pair production is determined from the
Bogolubov transformations. At high frequencies $\omega$ this rate
behaves as \ba \rho(\omega)|\gamma^{in}_{\vec{p}}|^2 \sim
e^{\omega/T_{H}}e^{-2\pi\omega}. \ea  Recalling the value of the
Hagedorn temperature $T_{H}=\f{1}{4\pi}$,  we see that this
expression diverges exponentially. The closed string pair
production is much more rapid than in the corresponding open
string case \cite{St}, where the exponentials cancelled and the
divergence was at most power law.

\section{The Two Point Function}
\setcounter{equation}{0} \hspace{5mm} In this section we define
the two point function in timelike Liouville theory by  analytic
continuation from the spacelike Liouville theory.

The two point function in the space-like theory is given by
\cite{Dorn:1992at}\cite{ZZ}\ba D(\ap)\equiv \la
e^{2\ap\phi}e^{2\ap\phi}\lb = (\pi\mu\gamma(b^2))^{\f{Q-2\ap}{b}}
\cdot \f{\gamma(2b\ap-b^2)}{b^2\cdot
\gamma(2-\f{2\ap}{b}+\f{1}{b^2})}, \ea where
$\gamma(x)=\Gamma(x)/\Gamma(1-x)$. This is related to the
reflection coefficient \ba &&\psi(\phi)\sim
e^{2ip\phi}+d(2ip)e^{-2ip\phi},\no &&d(2ip)\equiv
D(ip+\f{Q}{2})=-(\pi\mu \gamma(b^2))^{-2ip/b}
\cdot\f{\Gamma(1+2ip/b)\Gamma(1+2ipb)}{\Gamma(1-2ip/b)\Gamma(1-2ipb)}.
\no
\ea

We wish to analytically continue to $\phi=-iX_0$ and
$p=-i\omega/2$. We further take $b\equiv i\beta$ to be purely
imaginary since we are ultimately interested in the bulk time-like
Liouville theory with a real linear dilaton. The reflection
coefficient for bulk time-like Liouville theory is then \ba
d(\omega)=\la e^{-i(\omega +Q)X^0}e^{-i(\omega+Q)X^0} \lb =
-(\pi\mu\gamma(-\beta^2))^{i\f{\omega}{\beta}}
\cdot\f{\Gamma(1-i\omega/\beta)\Gamma(1+i\omega\beta)}
{\Gamma(1+i\omega/\beta)\Gamma(1-i\omega\beta)}. \no \ea Its
magnitude, which appears in the particle creation rate,  is simply
\ba |d(\omega)|=e^{-\f{\pi\omega}{\beta}}. \ea  This agrees
exactly with the $\beta=1$ minisuperspace result (\ref{tpf}).
Taking the limit $\beta\to 1$ 
with $\beta^2=1-\epsilon$ and
$\epsilon\to 0+$, we find  \ba d(\omega)\to
-(\pi\mu_{R})^{i\omega}\cdot e^{-\pi \omega}. \label{rrr}\ea where
we have identified the renormalized coupling
$\mu_{R}=|\mu\gamma(\beta^2)|$ from the expected invariance of the
two point function (for $Q=0$) under $X^0\to X^0+C,~~~ \mu_R \to
e^{2i\omega C}\mu_R$.  Note that we must take $\mu \to 0$ to keep
$\mu_R$ finite in the limit $\beta\to 1$. We will see in the next
section that this renormalization procedure also removes a similar
divergence in the three (and presumably higher) point function.

\section{The Three Point Function}
\setcounter{equation}{0} \hspace{5mm} In this section we use the
same  analytic continuation to  obtain a candidate three point
functions for  timelike Liouville theory. However subtleties are
encountered along the way. The result has some puzzling behavior
and may not be the correct three-point function.

 The three point
function in the spacelike theory is given by \cite{Dorn:1992at,ZZ}
(we use the conventions of the latter reference)\ba &&
C(\ap_1,\ap_2,\ap_3)\equiv \la
e^{2\ap_1\phi}e^{2\ap_2\phi}e^{2\ap_3\phi}\lb =[\pi
\mu\gamma(b^2)b^{2-2b^2}]^{(Q-\ap_1-\ap_2-\ap_3)/b}\times \no &&
\f{\Upsilon_0\Upsilon(2\ap_1)\Upsilon(2\ap_2)\Upsilon(2\ap_3)}
{\Upsilon(\ap_1+\ap_2+\ap_3-Q)
\Upsilon(\ap_1+\ap_2-\ap_3)\Upsilon(\ap_2+\ap_3-\ap_1)
\Upsilon(\ap_1+\ap_3-\ap_2)},\no \label{csl} \ea where
$\Upsilon_0=\f{d\Upsilon(x)}{dx}|_{x=0}$ and the function
$\Upsilon(x)$ is defined by \ba
\log\Upsilon(x)=\int^{\infty}_0\f{dt}{t}\left[(Q/2-x)^2 e^{-2t}
-\f{\sinh^2(Q/2-x)t}{\sinh(bt)\sinh(t/b)}\right]. \label{up} \ea

The analytic continuation of the prefactor in (\ref{csl}) to
$b=i\beta$ is given by \ba \left[e^{\pi i(\beta^2-1)}\cdot
(\pi\mu_R\beta^{2(1+\beta^2)})^{\f{i(\ap_1+\ap_2+\ap_3)}{\beta}}
\right]\!\cdot\! (\pi\mu_R\beta^{2(1+\beta^2)})^{1-1/\beta^2}\!\! \cdot\!
e^{-\pi\beta(\ap_1+\ap_2+\ap_3)}\!, \ea where all of the phase
factors are in the first parenthesis. Taking the limit $\beta\to
1$ yields the following expression for the prefactor \ba
(\pi\mu_{R})^{i(\ap_1+\ap_2+\ap_3)}\cdot
e^{-\pi(\ap_1+\ap_2+\ap_3)}. \ea

The non-trivial part is the analytic continuation of the function
$\Upsilon(x)$ appearing in (\ref{csl}), which acquires an infinite
accumulation of poles for purely imaginary $b$. This difficulty
was already encountered at the level of the two-point function for
timelike boundary Liouville theory, and our mathematical treatment
will follow the one given in \cite{GS}.  We would like to perform
an analytical continuation from the real value of $b$ to the
imaginary value of $b=i\beta$ by a $\pi/2$ rotation. In deforming
the contour of integration of (\ref{up}) along the real axis into
one along the imaginary axis, we encounter poles at $t=\f{n\pi
i}{b}\ \ (n=1,2,\ddd)$. Hence one may write the integral over the
real axis as an integral $I$ along the imaginary axis plus a sum
of poles $P$ \ba \log\Upsilon(x)=I(x)+P(x). \ea

The convergent integral $I(x)$ along the imaginary axis $t=i\tau$
is given by \ba I(x)=\int^{\infty}_{0}
\f{d\tau}{\tau}\Bigl[(Q/2-x)^2e^{-2i\tau}
-\f{\sin^2((Q/2-x)\tau)}{\sin(b\tau)\sin(\tau/b)}\Bigr]\no
=\int^{\infty}_{0} \f{d\tau}{\tau}\Bigl[(iq/2-x)^2e^{-2i\tau}
-\f{\sin^2((iq/2-x)\tau)}{\sinh(\beta
\tau)\sinh(\tau/\beta)}\Bigr], \label{ix1} \ea where we set
$b=i\beta$ and $Q=i(\beta-1/\beta)=iq$. It follows  from
(\ref{csl}) that we are interested in the linear combination
 \ba
I(\ap_1,\ap_2,\ap_3)&=&-I(\ap_1+\ap_2+\ap_3)-I(\ap_1+\ap_2-\ap_3)
-I(\ap_1-\ap_2+\ap_3)\no &&-I(-\ap_1+\ap_2+\ap_3) \ \
+I(2\ap_1)+I(2\ap_2)+I(2\ap_3).\label{dec} \ea in which the linear
and quadratic terms with respect to the argument $x$ cancel. Hence
we can omit\footnote{This leads to the divergence due to the pole
$\tau=0$ in (\ref{ix1}). However, it is cancelled in the ratio
(\ref{csl}).} the first term in (\ref{ix1}). Setting $\beta=1$ one
finds \ba
I(x)&=&-\int^{\infty}_{0}\f{d\tau}{\tau}\f{\sin^2(x\tau)}{\sinh^2\tau}\no
&=&\f{1}{2}\int^{\infty}_{0}\f{d\tau}{\tau}\f{\cos(2x\tau)}{\sinh^2\tau}
+const.\label{ix2} \ea To derive this we neglect the imaginary
part of $I$ since it is proportional to $x^2$. Note that
$e^{I(x)}$ has no zero or poles for real $x$. However, for e.g.
$x=\pm i\beta$, it has a zero.

The sum of poles is given by \ba
P(x)&=&-2i\sum_{n=1}^{\infty}\f{(-1)^n \sin^2(\f{\pi
n}{b}(Q/2-x))} {n\sin(\f{\pi n}{b^2})}\no
&=&-i\sum_{n=1}^{\infty}\f{(-1)^n}{n\sin(\f{\pi n}{b^2})}
+i\sum_{n=1}^{\infty}\f{1}{n}\Biggl(\f{\cos(\f{\pi n}{b^2})
\cos(\f{2\pi nx}{b})}{\sin(\f{\pi n}{b^2})}+\sin(\f{2\pi
nx}{b})\Biggr).\no \ea The first term is constant and can be
neglected. We drop this term and analytically continue defining
$\f{1}{b^2}=-\f{1}{\beta^2}-i\ep$ and $x=-iby$ and taking $y$ to
be real. The real part $\mbox{Re}P$ of $P(x)$ is
\ba
\mbox{Re}P&=&\sum_{n=1}^{\infty}\f{1}{n}\Biggl(\sinh(2\pi ny)+
\mbox{Im}\Bigl(\cot(\f{n\pi}{\beta^2}+in\pi\ep)\Bigr)\cosh(2\pi
ny) \Biggr). \ \ \ \ \label{real1} \ea The first term can be resummed by
using the
 analytic continuation\footnote{
 This follows from
\ba &&\sum_{n=1}^{\infty}\f{\sin(nz)}{n}=(\pi-z)/2,\no
&&\sum_{n=1}^{\infty}\f{\cos(nz)}{n}=-\log(2\sin(z/2)), \ea by
substituting $z=2\pi i y$.} \ba \sum_{n=1}^{\infty}\f{1}{2n}e^{2\pi
ny} -\sum_{n=1}^{\infty}\f{1}{2n}e^{-2\pi ny}
=-\f{1}{2}\log(1-e^{2\pi y})+\f{1}{2}\log(1-e^{-2\pi y})=-\pi
y+\f{\pi i}{2}. \no \ea These terms cancel one another in the ratio
(\ref{csl}) and can be neglected 
except that it leads to a constant factor $-\pi$
in $\Upsilon_{0}$. 
The second term in (\ref{real1}) can be rewritten as
\ba -\f{1}{2}\sum_{n=1}^{\infty} \f{\cosh(2\pi ny)}{n}
\cdot\f{\sinh(2\pi n\ep)}{\sinh^2(\pi
n\ep)+\sin^2(\f{n\pi}{\beta^2})}. \ea Assuming $\beta^2$ is
irrational as in \cite{GS} we find that this contribution vanishes
in the limit $\ep\to 0$.

The imaginary part of the pole sum is
 \ba
\mbox{Im}P&=&-\sum_{n=1}^{\infty}\f{1}{n}
\mbox{Re}\Bigl(\cot(\f{n\pi}{\beta^2}+in\pi\ep)\Bigr)\cosh(2\pi
ny). \ea If the $\beta^2$ is rational such that $\beta^2=q/p$,
then \ba
\mbox{Re}\Bigl(\cot(\f{n\pi}{\beta^2}+in\pi\ep)\Bigr)=\f{1}{2}
\f{\sin(2np\pi/q)}{\sinh^2(\pi n\ep)+\sin^2(\f{np\pi}{q})}. \ea
The second term can also be rewritten after we take the limit
$\ep\to 0$ \ba -\sum_{n_0=1}^{q-1}\sum_{m=0}^{\infty}\f{\cosh(2\pi
ny)}{mq+n_0} \cot(n_0\pi\f{p}{q}). \ea
For the specific value $\beta=1$ this contribution also vanishes.\\

Putting this all together leads to the very simple formula for the
three-point function at $\beta=1$ ($Q=0$) \ba
C(\omega_1,\omega_2,\omega_3)&\equiv& 
\la e^{-i\omega_1X^0}e^{-i\omega_1X^0}e^{-i\omega_1X^0} \lb \no
&=&-\pi(\pi\mu_{R})^{i(\omega_1+\omega_2+\omega_3)}\cdot
e^{-\pi(\omega_1+\omega_2+\omega_3)/2}
\cdot e^{I(\omega_1/2,\omega_2/2,\omega_3/2)}. 
\no \label{vbn}\ea As a check
we note that (\ref{vbn}) reduces to the two-point
function (\ref{rrr}) up to a constant when we set $\omega_3=0$ and
$\omega_1=\omega_2=\omega$.

However the candidate three point function has a disturbing
feature which leads us to suspect its validity. Conformal
invariance would lead us to expect that it should vanish for
$\omega_3=0$ and $\omega_1\neq \omega_2$, but it does not do so.
It can be seen that the original three point function (\ref{csl})
indeed vanishes for such momenta when Re$b ~>0$.

How is it possible that such correlators vanish for all $\ep \neq
0$, but not at $\ep =0$? The answer lies in the order of limits
used in defining the continued expressions. Suppose that we do not
take the limit $\ep\to 0$ before $n\to \infty$ in (\ref{real1}).
In this case we can write  \ba &&\sum_{n=1}^{\infty}\f{1}{n}
\mbox{Im}\Biggl(\cot(\f{n\pi}{\beta^2}+in\pi\ep)\Biggr)\cosh(2\pi
ny)\no &&\ \ \ =\sum_{n=1}^{\infty}\f{1}{n}\mbox{Re}
\Biggl(\f{e^{in\pi/\beta^2-n\pi\ep}+e^{-in\pi/\beta^2+n\pi\ep}}
{e^{in\pi/\beta^2-n\pi\ep}-e^{-in\pi/\beta^2+n\pi\ep}}\Biggr)
\cosh(2\pi ny).\label{alm} \ea Since we can replace the fraction
with -1 for large $n$ and $\ep>0$, 
we expect that the sum at $y=0$ is divergent
even though the value at non-zero $y$ is zero in the $\ep\to 0$
limit. Thus we could have
$\mbox{Re}P\to -\infty$ at $y=0$,
producing the expected zero in the three-point correlator.

This raises the possibility that the mathematical continuation
procedure used to compute the three point correlator (\ref{vbn})
is not the physically correct one.  However we have not been able
to find an alternate procedure which yields correlators that are
analytic functions of the momenta and vanish at the expected
places. We hope that future work will either provide an
explanation of the unusual behavior of (\ref{vbn}) or an alternate
expression.

\bigskip
\begin{center}
\noindent{\large \bf Acknowledgments}
\end{center}
We are grateful to M. Gutperle, S. Minwalla, L. Motl and N. Toumbas 
for useful conversations.  This
work was supported in part by DOE grant DE-FG02-91ER40654.

\end{document}